\documentclass[12pt]{article}
\usepackage{graphicx}
\def \app{D_{\pi \pi}}

\def \bea{\begin{eqnarray}}
\def \beq{\begin{equation}}

\def \bo{B^0}

\def \alef{\alpha_{\rm eff}}

\def \cn{Collaboration}
\def \cpp{C_{\pi \pi}}
\def \crr{C^{\rm long}_{\rho \rho}}
\def \eea{\end{eqnarray}}
\def \eeq{\end{equation}}
\def \ite{{\it et al.}}

\def \lpp{\lambda_{\pi \pi}}
\def \ob{\overline{B}^0}

\def \rpp{R_{\pi \pi}}

\def \spp{S_{\pi \pi}}
\def \srr{S^{\rm long}_{\rho \rho}}
\begin{document}
\begin{flushright}
TECHNION-PH-2004-44\\
ZU-TH-16-04\\
hep-ph/0410170 \\
October 2004 \\
\end{flushright}
\renewcommand{\thesection}{\Roman{section}}
\renewcommand{\thetable}{\Roman{table}}
\centerline{\bf REDUCING THE ERROR ON $\alpha$ in 
$B\to \pi\pi,~\rho\rho,~\rho\pi$}
\medskip
\centerline{Michael Gronau}
\centerline{\it Physics Department, Technion -- Israel Institute of 
Technology}
\centerline{\it 32000 Haifa, Israel}
\medskip
\centerline{Enrico Lunghi and Daniel Wyler}
\centerline{\it Institut f\"ur Theoretische Physik, Universit\"at Z\"urich,
Winterthurerstrasse 190}
\centerline{\it CH-8057, Z\"urich, Switzerland}
\bigskip
\begin{quote}
Theoretical errors in the extraction of  $\alpha$ from $B\to
(\pi^+\pi^-,~\rho^+\rho^-,~\rho^{\pm}\pi^{\mp})$ decays are usually
given in terms of upper bounds on $|\alef - \alpha|$ obtained from
isospin or from SU(3) relations, where $\alef$ is measured through CP
asymmetries. We show that mild assumptions about magnitudes and strong
phases of penguin and tree amplitudes ($|P/T| \le1$ and $|\delta| \le
\pi/2$) in $B\to\pi\pi$ and $B\to\rho\rho$, imply $\alef > \alpha$,
thus reducing by a factor two the error in $\alpha$. Similarly, the
assumptions $|p_{\pm}/t_{\pm}| \le 1,~|\delta_-| \le \pi/2 \le
|\delta_+|$ in $B\to\rho\pi$ lead to a cancellation between two terms
in $\alef -\alpha$.  Current data support these conditions,
%MG added
which are justified by both QCD-factorization and flavor SU(3).
\end{quote}

\leftline{\qquad PACS codes:  12.15.Hh, 12.15.Ji, 13.25.Hw, 14.40.Nd}

\bigskip
Direct extraction of the Cabibbo-Kobayashi-Maskawa (CKM) phase
$\alpha\equiv \phi_2$ from the time-dependent CP asymmetry in 
$B^0\to\pi^+\pi^-$ is obstructed by the penguin amplitude~\cite{Pen}. 
This obstacle may be
overcome using isospin symmetry~\cite{iso}, which incorporates
electroweak penguin contributions at a percent level~\cite{EWP} but
does not include small isospin breaking 
effects~\cite{Gardner}. This method requires
separate rate measurements of $\bo,\ob \to \pi^0\pi^0$. As long as low
statistics does not permit these separate measurements, one can use
combined $B$ and $\overline B$ decay rates into the three $\pi\pi$
channels to obtain upper bounds on $|\alef^{\pi\pi}-
\alpha|$~\cite{GQ,Ch,GLSS}. The angle $\alef^{\pi\pi}$, which equals
$\alpha$ in the limit of a vanishing penguin amplitude, is given up to
a discrete ambiguity by the time-dependent CP asymmetry in
$B^0\to\pi^+\pi^-$. The upper bound on $|\alef^{\pi\pi}- \alpha|$,
improved by information from an upper limit on the asymmetry in
$B\to\pi^0\pi^0$, remains an intrinsic theoretical uncertainty in
$\alpha$. Note that $\alef^{\pi\pi}- \alpha$ may be either positive or
negative.

Similar considerations can also be applied to $B^0\to\rho^+\rho^-$ and
lead to analogous bounds on $|\alef^{\rho\rho} - \alpha|$. Because
each of the two $\rho$ mesons carries a unit spin, one must
distinguish between decays to even-CP and odd-CP final states
corresponding to definite polarizations~\cite{DQSTL}.  

%MG On the other hand, 
A complete isospin analysis of the processes
$B\to\rho\pi$ is complicated by the existence of five different
$\rho\pi$ charge states in $B^0$ and $B^+$ decays~\cite{rhopi}. Thus,
it was proposed to measure $\alpha$ through the time-dependent Dalitz
distribution in $B^0\to \pi^+\pi^-\pi^0$~\cite{SQ}, which provides
information about interference of amplitudes for $B^0\to \rho^+\pi^-,
B^0\to \rho^-\pi^+$ and $B^0\to \rho^0\pi^0$. Alternatively, one may
measure an angle $\alef^{\rho\pi}$ in quasi two- body decays
$B^0(t)\to \rho^{\pm}\pi^{\mp}$, and use broken flavor SU(3) to obtain
an upper bound on $|\alef^{\rho\pi} - \alpha|$~\cite{GZ}.
   
The experimental progress during the past year has been impressive in
this class of measurements. The situation in early July~\cite{Beach}
was updated in late August~\cite{ZL}, following new measurements
reported at the International Conference on High Energy Physics in
Beijing~\cite{Giorgi}. 
The overall range of $\alpha$ determined in $B\to
\pi\pi,~\rho\rho,~\rho\pi$~\cite{Giorgi},  $\alpha = (100^{+12}_{-11})^\circ$, 
overlaps with and begins to be narrower than the following bounds 
obtained indirectly in an independent global CKM fit~\cite{Charles},
\beq\label{CKM} 
78^\circ \le \alpha \le 122^\circ~,~~~~~~ 38^\circ \le
\gamma \le 80^\circ~, 
\eeq 
where a 95$\%$ confidence level (CL) is implied.  The error in $\alpha$ from a
time-dependent Dalitz plot analysis of $B \to
\pi^+\pi^-\pi^0$~\cite{Giorgi}, $(^{+27}_{-17}\pm 6)^\circ$, is
statistics-dominated. On the other hand, the theoretical errors in
determining $\alpha$, using isospin in $B\to\pi\pi,~\rho\rho$ and
applying broken SU(3) to $B\to\rho\pi$, are at least as large as the
corresponding statistical errors. The 90$\%$ CL upper bounds on
$|\alpha_{\rm eff} - \alpha|$ in these three cases 
are~\cite{GZ,Giorgi,Hocker} $37^\circ,15^\circ$ and 
%MG correcterd for SU(3) breaking 15 to 17=13x1.3
$17^\circ$,
respectively. It would be very useful to reduce these intrinsic
theoretical uncertainties using present data.

In this letter we point out that the above theoretical errors in
$\alpha$ may be reduced by about a factor two under very mild and
reasonable assumptions,
%MG added
which are justified by both QCD-factorization and flavor SU(3).
%MG changed sentence
In $B\to\pi^+\pi^-$ and $B\to\rho^+\rho^-$ we study the dependence of 
$\alef - \alpha$ on the ratio of penguin and tree amplitudes, $r\equiv |P/T|$, 
and on their relative strong phase, $\delta$. We show that the conditions
$|P/T| \le 1$ and $|\delta| \le \pi/2$ predict a positive sign for
$\alpha_{\rm eff} -\alpha$.
%MG changed sentence
In $B\to \rho^{\pm}\pi^{\mp}$ two small ratios of penguin and tree 
amplitudes, $r_{\pm}$, and two phases $\delta_{\pm}$ lying in opposite 
hemispheres, tend to suppress $\alpha_{\rm eff} -\alpha$ due
to a cancellation between two terms. We propose to include, in future
studies of $\alpha$, the explicit dependence of $\alef - \alpha$ on
$r$, $\delta$, $r_\pm$ and $\delta_\pm$, together with the dependence
of the CP asymmetries on these variables. In the argumentation below
we assume $\gamma< 90^\circ$ as given in~(\ref{CKM}).

To prove our point, we consider first $B^0\to\pi^+\pi^-$. We use the
$c$-convention defined in~\cite{GRconv}, in which the top-quark has
been integrated out in the $b\to d$ penguin transition and unitarity
of the CKM matrix has been used. Absorbing a $P_{tu}$ term in $T$, the
decay amplitude may be written in the following general form,
\beq\label{amp}
A(B^0\to\pi^+\pi^-) = T + P = |T|\left (e^{i\gamma} + re^{i\delta}\right )~,
\eeq
where by convention $r > 0,~-\pi < \delta \le \pi$. 
The phase $\alpha^{\pi\pi}_{\rm eff}$ is extracted up to a discrete
ambiguity from the two asymmetries $\spp$ and $\cpp$ in
$B^0(t)\to\pi^+\pi^-$~\cite{MG}:
\beq\label{alefSC}
\sin(2\alpha^{\pi\pi}_{\rm eff}) = \frac{\spp}{\sqrt{1 - \cpp^2}}~,
\eeq 
where
\bea\label{cpp}
\cpp & \equiv & \frac{1 - |\lpp|^2}{1 + |\lpp|^2}
= \frac{2r\sin\delta\sin\gamma}{\rpp}~,\\
\label{spp}
\spp & \equiv & \frac{2 {\rm Im}(\lpp)}{1 + |\lpp| ^2}
=\frac{\sin 2\alpha - 2r\cos\delta\sin(\alpha - \beta) - 
r^2\sin 2\beta}{\rpp}~,\\
\rpp &\equiv& 1 + 2r\cos\delta\cos\gamma + r^2~.
\eea
Using the definitions
\beq\label{def-alef}
\alef^{\pi\pi}\equiv \frac{1}{2}{\rm Arg}\lpp~,~~~~~
\lpp \equiv e^{-2i \beta} \frac{A(\ob \to \pi^+ \pi^-)}
{A(B^0 \to \pi^+ \pi^-)}~,
\eeq
one has
\beq\label{al1}
\alef^{\pi\pi} - \alpha = \frac{1}{2}\arctan\left [
\frac{2r\sin\gamma(\cos\delta + r\cos\gamma)}
{1 - r^2 + 2r\cos\gamma(\cos\delta + r\cos\gamma)}\right ]~.
\eeq
Two very reasonable assumptions, $r \le 1$ and $\cos\delta >-
r\cos\gamma$, imply $\alef^{\pi\pi} - \alpha > 0$. The second condition 
may be replaced by a stronger one, $\cos\delta \ge 0$; it is stronger since
we are assuming $\gamma< 90^\circ$. The result $\alef^{\pi\pi} >
\alpha$ demonstrates the central point of this note.

Rewriting Eq.~(\ref{al1}),
\beq\label{al2}
\cot2(\alef^{\pi\pi} - \alpha) = \cot\gamma + \frac{1-r^2}
{2r\sin\gamma(\cos\delta  + r\cos\gamma)}~,
\eeq
we see that the above assumptions imply $\cot 2(\alef^{\pi\pi} -
\alpha) > \cot\gamma$, or $0 < \alef^{\pi\pi} - \alpha < \gamma_{\rm
max}/2 = 40^\circ$. The prediction of the sign of $\alef^{\pi\pi}
-\alpha$ is based purely on definitions and on the assumptions on $r$
and $\delta$. It implies a considerably narrower range for $\alpha$
than the bound $|\alef^{\pi\pi} - \alpha| <
37^\circ$~\cite{Giorgi,Hocker}, obtained from the $B\to\pi\pi$ and
$\bar B\to \pi\pi$ isospin triangles, with the inclusion of some
information about the asymmetry in $B\to\pi^0\pi^0$.

It now remains to justify our two assumptions, $r \le 1$ and $|\delta|
\le \pi/2$.  The ratio $r$ has a very long history, starting in the
late eighties when the penguin amplitude in $B \to \pi^+\pi^-$ was
estimated to be small but non-negligible~\cite{Pen,PenGr}. First
measurements of $B\to \pi\pi$ and $B \to K\pi$ decay rates, performed
several years later by the CLEO collaboration~\cite{CLEO}, were
analyzed within flavor SU(3) indicating that $r \sim
0.3$~\cite{DGR}. A recent global SU(3) fit to all $B \to \pi\pi$ and
$B\to K\pi$ decays obtained an unexpected large value~\cite{CGRS} $r =
0.69 \pm 0.09$. The large value of $r$ is driven partly but not
only~\cite{GRpipi} by a large input value for
$|\cpp|$~\cite{BFRS,ALP,BPRS}. Theoretical calculations based on QCD
and a heavy quark expansion~\cite{BBNS,KLS} find somewhat smaller
values for $r$, all lying comfortably in the range $r<0.5$.  All these
calculations support strongly the assumption $r \le 1$.

Very early theoretical QCD arguments favoring a small value of
$|\delta|$ were proposed in~\cite{BSS}. Although different
calculations of $\delta$~\cite{BBNS,KLS}, based on QCD and a heavy
quark expansion, do not always agree in detail, all these computed
values of $|\delta|$ are considerably smaller than $90^\circ$.
Constraints on CKM parameters based on the assumption $|\delta|\le
90^\circ$ and on a given range for $\spp$ were studied
in~\cite{BS}. Long distance $c\bar c$ penguin
contributions~\cite{CFMS}, or equivalently final state
rescattering~\cite{BSV,CCS}, may spoil the QCD
calculations~\cite{BPRS}. These effects cannot be calculated in a
model-independent way. A global SU(3) fit to $B \to \pi\pi$ and $B\to
K\pi$ decays, which effectively includes these rescattering effects
while assuming SU(3) invariant strong phases, obtains~\cite{CGRS}
$\delta = (-34^{+11}_{-25})^\circ$.  Large SU(3) breaking effects in
strong phases could possibly lead to values of $\delta$ outside the
range $|\delta| \le \pi/2$. 

A study of the two hadronic parameters, $r$ and $\delta$, was
performed in~\cite{ALP}, adding measurements of $\spp, \cpp$ and an
upper bound on $|\alef^{\pi\pi} - \alpha|$ available before this
summer to all other CKM constraints~\cite{Charles}. Values were found
in an overall minimum $\chi^2$ fit, $r = 0.77 ^{+0.58}_{-0.34}$ and
$\delta = (-43^{+14}_{-21})^\circ$, consistent with $|\delta| \le
90^\circ$ but possibly violating somewhat the bound $r\le 1$. A recent
update~\cite{Ali}, using newer measurements, also favors parameters in
the range $r\le1,~-90^\circ \le \delta\le 0^\circ$.

Similarly, we attempt a direct experimental proof of $r \le 1$ and $|\delta| \le
\pi/2$ using current data~\cite{Giorgi,HFAG}:
\beq\label{SCpipi}
\spp = -0.61 \pm 0.14~(0.32)~,~~\cpp = -0.37 \pm 0.11~(0.27)~,~
|\alef^{\pi\pi} - \alpha|< 37^\circ~(90\%~{\rm CL})~,
\eeq
The world averaged asymmetries are based on most recent measurements 
by the Belle collaboration~\cite{Belle} ($\spp = -1.00\pm 0.21\pm 0.07,
\cpp = -0.58\pm 0.15\pm 0.07$) and by the BaBar collaboration~\cite{BaBar} 
($\spp = -0.30\pm 0.17\pm 0.03, \cpp = -0.09\pm 0.15\pm 0.04$). 
Since these two measurements are not in good agreement with each other, 
an error rescaling factor of 2.39~\cite{Hocker} (determining errors in parentheses) 
may be used to achieve a conservative confidence level.  Expressions for
$\cpp,~\spp$, and $|\alef^{\pi\pi} - \alpha|$, given in Eqs.~(\ref{cpp})(\ref{spp}) and 
(\ref{al1}), and the values in Eq.~(\ref{SCpipi}) can be used to constrain $r, \delta$ 
and $\alpha$. 

\begin{figure}
\includegraphics[height=2.95in]{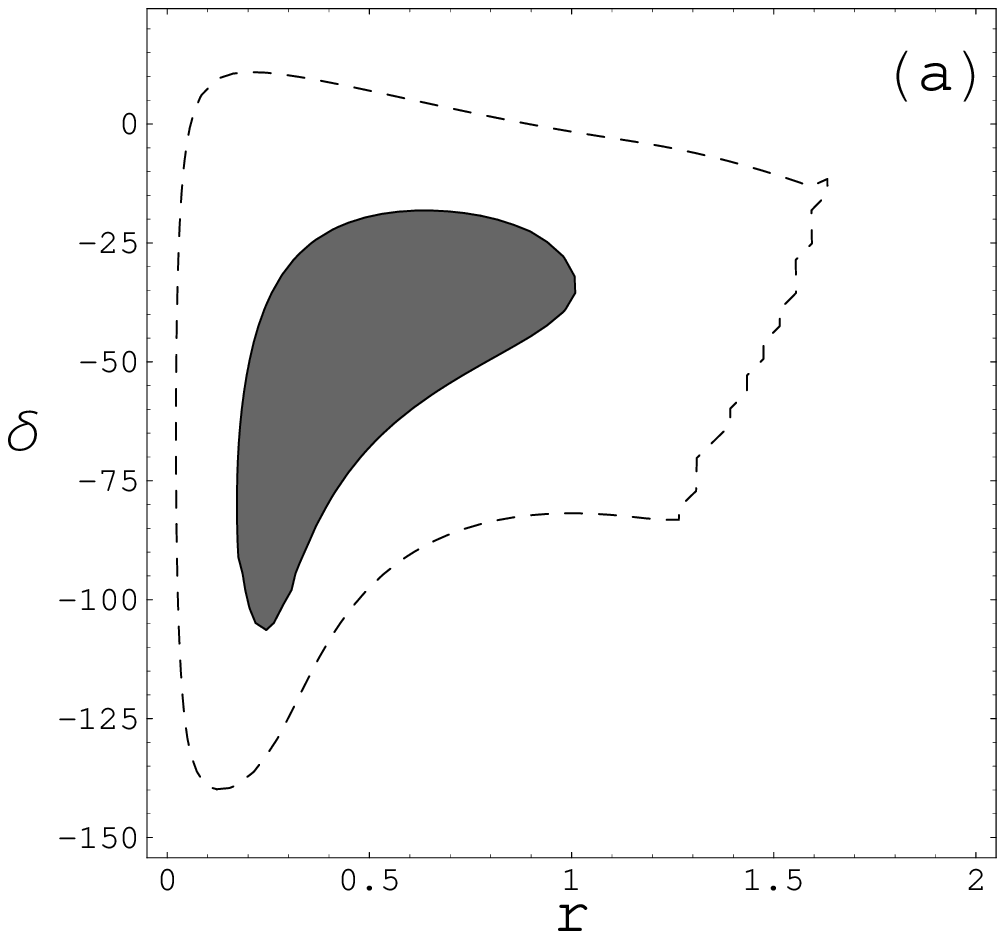}
\includegraphics[height=2.95in]{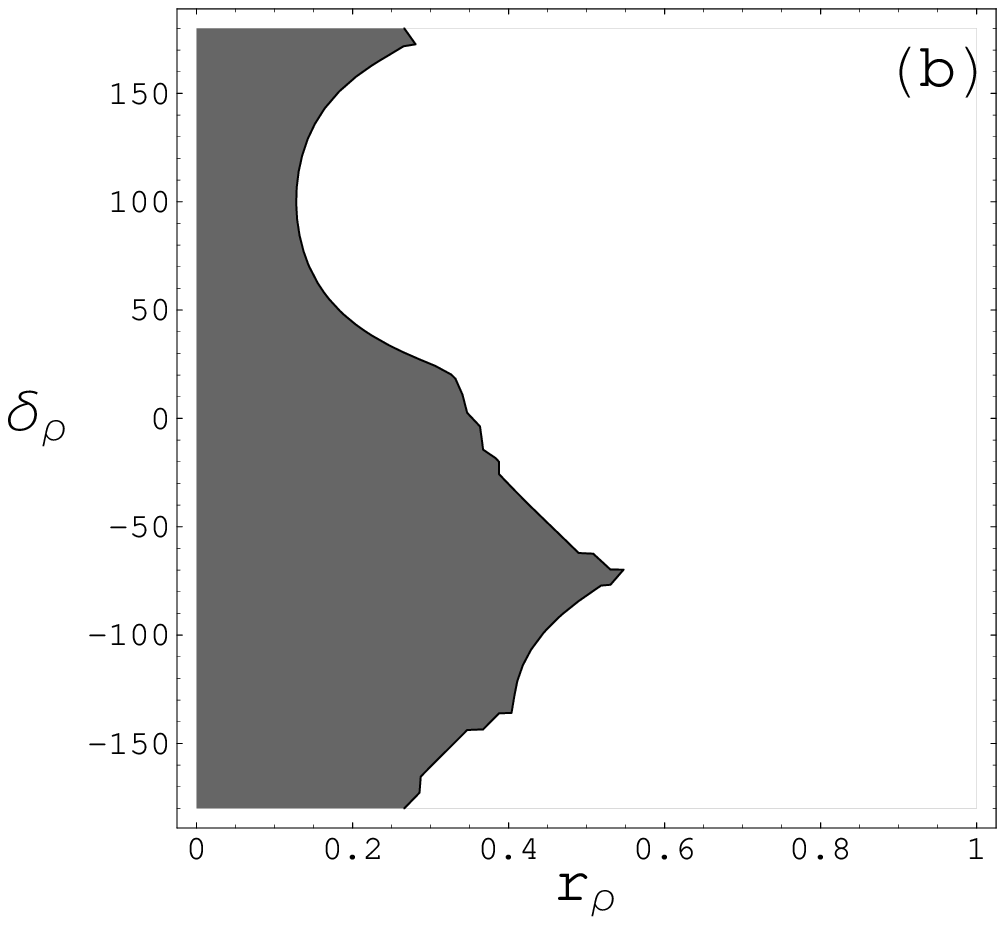}
\caption{{\bf (a)} Bounds on $r$ and $\delta$ at 90$\%$ CL based on
a $\chi^2$ analysis of Eq.~(\ref{SCpipi}). The shaded and dashed areas
correspond to unscaled and rescaled errors in (\ref{SCpipi}),
respectively. {\bf (b)} 90$\%$ CL bound on $r_\rho$ and
$\delta_\rho$ based on (\ref{SCrr}).}
\label{fig_corr}
\end{figure}

We performed a separate minimum-$\chi^2$ fit for the parameters $r$
and $\delta$. The $\chi^2$ contains two contributions from $\cpp$ and
$\spp$ and the usual terms corresponding to the standard analysis of
the unitarity triangle. In the minimization we reject points which
yield $|\alef^{\pi\pi} - \alpha| \ge 37^\circ$. 
Fig.~\ref{fig_corr}a shows the resulting 90$\%$ CL bounds obtained
using the unscaled (dark area) and rescaled (dashed line) errors in
Eq.~(\ref{SCpipi}). The bounds can be summarized by
\bea
& 0.2~(0.05)  <  r  < 1~(1.65)~,\nonumber\\
&  -107^\circ~(-140^\circ) < \delta < -19^\circ~(10^\circ)~,
\eea
where the numbers in parentheses correspond to the rescaled errors.
Note that the unscaled errors exclude $r\ge 1$ at 90$\%$ CL,
favoring $-90^\circ \le \delta \le 0$. Although they do not exclude
completely $\delta < -90^\circ$, we find that the lowest allowed value
of $\alef^{\pi\pi} -\alpha$ is $-3^\circ$, quite close to zero. The
rescaled errors imply less restrictive bounds. Exclusion of
$r>1,~\delta < -90^\circ$ would imply $\alpha < \alef^{\pi\pi} =
110.5^\circ$ for the central values given in Eq.~(\ref{SCpipi}). In
any case, using the dependence of $\alef^{\pi\pi} - \alpha$, $\spp$ and
$\cpp$ on $r$ and $\delta$ is expected to reduce the theoretical
uncertainty in $\alpha$ below the upper limit on $|\alef^{\pi\pi} -
\alpha|$.
\begin{figure}
\begin{center}
\includegraphics[height=3.5in]{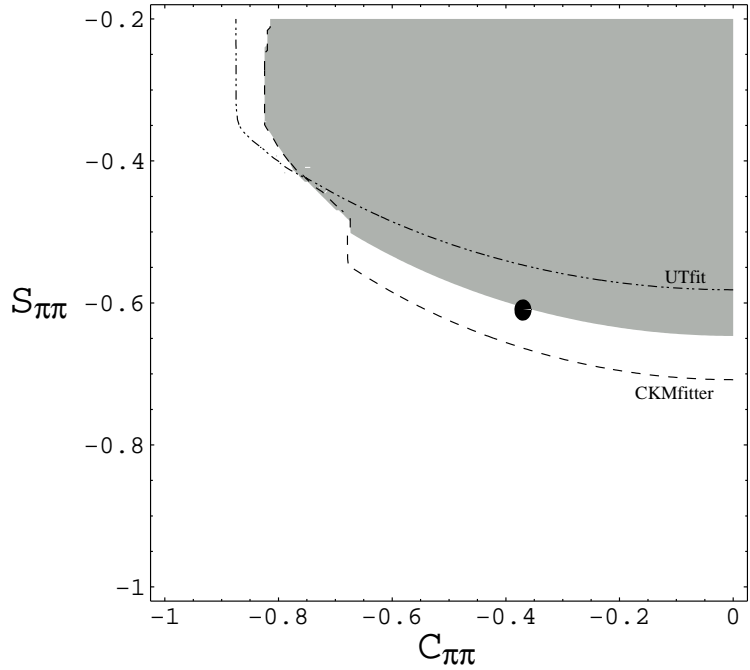}
\end{center}
\caption{Regions in the ($\cpp,\spp$) plane {\em outside} of which
$\alef^{\pi\pi}-\alpha>0$ at 90$\%$ CL. The shaded area is obtained
using the UT analysis presented in Ref.~\cite{ALP}. The dashed and
dotted lines correspond to analyses of the unitarity triangle by the 
CKMfitter~\cite{Charles} and
UTfit~\cite{Bona} collaborations.
The heavy dot denotes the current central values (\ref{SCpipi}).}
\label{fig_CS}
\end{figure}

One may study directly the 
dependence of the sign of $\alef^{\pi\pi}-\alpha$ on the
values of $\cpp$ and $\spp$. Assuming that the discrepancy between 
the BaBar and Belle  results disappears and taking errors
$\delta \cpp \sim \delta \spp \sim 0.1$, we calculate central values in the 
($\cpp,\spp$) plane for which $\alef^{\pi\pi}-\alpha>0$ at 90$\%$ CL.
An important input is our prior knowledge of $\gamma$. 
For this we use three different analyses of the unitarity triangle described 
in Refs.~\cite{ALP},~\cite{Charles} and~\cite{Bona}. Corresponding  
regions in the ($\cpp,\spp$) plane, {\em outside} of which
$\alef^{\pi\pi}-\alpha>0$, are described in Fig.~\ref{fig_CS} by the shaded 
area and by the dashed and  dotted lines, respectively. The heavy dot, denoting 
the current central values of $\cpp$ and $\spp$,  is seen to lie close to the 
bounds but still {\em inside} the most conservative region~\cite{Charles}.

Let us now turn to discuss briefly $\alef^{\rho\rho} - \alpha$ in
$B\to \rho^+\rho^-$. Using the approximately pure longitudinal
polarization of the final states~\cite{Giorgi,Pol}, the isospin analysis
of the decays $B\to\rho\rho$ is greatly simplified, becoming almost
identical to that of $B\to\pi\pi$ (neglecting the small $\rho$-width
effect~\cite{FLNQ}). The small branching ratio for 
$B\to\rho^0\rho^0$~\cite{rho0rho0} leads to the rather tight 90$\%$ 
CL upper bound~\cite{Giorgi,Hocker}, $|\alef^{\rho\rho} - \alpha| <
15^\circ$. The phase difference $\alef^{\rho\rho} - \alpha$ may be
written as in Eq.~(\ref{al1}) and becomes positive under similar
conditions for $r_{\rho}$ and $\delta_{\rho}$. That is, our generic
assumptions, $r_{\rho} \le 1$ and $|\delta_{\rho}| \le \pi/2$
for longitudinally polarized $\rho$ mesons, imply $\alef^{\rho\rho} > \alpha$. 

Although these assumptions about $r_\rho$ and $\delta_\rho$ seem
reasonable~\cite{Aleksan}, one may attempt to verify them experimentally, as
demonstrated above for $B\to \pi\pi$.  Current measurements of $B(t)
\to\rho^+\rho^-$ are consistent with a zero CP
asymmetry~\cite{Giorgi},
\beq\label{SCrr}
\srr = -0.19 \pm 0.35~,~~~\crr = -0.23 \pm 0.28~,~~~|\alef^{\rho\rho} - 
\alpha| < 15^\circ~(90\%~{\rm CL})~.
\eeq 
The central values of $\srr$ and $\crr$ imply $\alef^{\rho\rho} = 95.6^\circ$.  
An analysis, based on expressions for $\crr,~\srr$ and
$\alef^{\rho\rho}-\alpha$ analogous to Eqs.~(\ref{cpp}) (\ref{spp}) and
(\ref{al1}), in terms $r_{\rho},~\delta_{\rho}$, and $\gamma$, proceeds 
as in $B\to\pi\pi$.  The rather low
upper bound, $|\alef^{\rho\rho} - \alpha| < 15^\circ$, implies a small
value for $r_\rho$, as can be seen in Eq.~(\ref{al1}) where
$\alef^{\rho\rho} - \alpha$ vanishes in the limit $r_\rho\to 0$. A
small value of $r_\rho$ means a small penguin amplitude and a 
low sensitivity to the value of $\delta_\rho$.

We have verified the above statement by performing a $\chi^2$ fit
based on Eq.~(\ref{SCrr}). In Fig.~\ref{fig_corr}b, we show the 90$\%$
CL for the parameters $r_\rho$ and $\delta_\rho$. We see that while
$r_\rho>1$ is excluded ($0 < r_\rho < 0.55$), values of $\delta_\rho$
are permitted in the entire range $-\pi < \delta_\rho \le \pi$. This
situation is unlikely to change with a reduction of errors in $\srr$
and $\crr$ if $r_\rho$ is much smaller than one
%MG 
as calculated in naive factorization~\cite{Aleksan}. 

The situation in $B\to\rho^{\pm}\pi^{\mp}$ is somewhat more
complicated than the one in $B\to\pi^+\pi^-$ and
$B\to(\rho^+\rho^-)_{\rm long}$ because the final states
$\rho^{\pm}\pi^{\mp}$ are not CP-eigenstates. Time-dependent decay
rates for initially $B^0$ decaying into $\rho^\pm\pi^\mp$ are given
by~\cite{MGPLB}:
\beq\label{Gammat}
\Gamma(B^0(t) \to \rho^\pm\pi^\mp) \propto  1 + (C \pm \Delta C)\cos\Delta mt
- (S \pm \Delta S)\sin\Delta mt~.
\eeq
For initially $\ob$ decays the $\cos\Delta mt$ and $\sin\Delta mt$
terms have opposite signs. One defines a measurable phase
$\alef^{\rho\pi}$~\cite{GZ}, which equals $\alpha$ in the limit of
vanishing penguin amplitudes,
\beq \label{alefrp}
\alpha^{\rho\pi}_{\rm eff} \equiv  \frac{1}{4}\left [\arcsin\left (\frac{S + \Delta S}
{\sqrt{1- (C + \Delta C)^2}}\right )
+ \arcsin\left (\frac{S - \Delta S}{\sqrt{1- (C - \Delta C)^2}}\right )\right ]~.
\eeq

The phase $\alef^{\rho\pi}$ can be expressed in terms of parameters
defining decay amplitudes. One denotes two distinct amplitudes for
$B^0\to\rho^+\pi^-$ and $B^0\to\rho^-\pi^+$ by the charge of the $\rho$ meson,
\bea\label{amprhopi}
A(B^0\to\rho^+\pi^-) & =  & |t_+|(e^{i\gamma} + r_+e^{i\delta_+})~,\\
A(B^0\to\rho^-\pi^+) & = & |t_-|(e^{i\gamma} + r_-e^{i\delta_-})~,
\eea
where $r_{\pm}\equiv |p_{\pm}/t_{\pm}|$ are two ratios of penguin and
tree amplitudes, and $-\pi < \delta_{\pm}\le \pi$ are the corresponding relative
strong phases. Continuing the analogy with $B\to \pi^+\pi^-$ by defining
\beq\label{def-alefrp}
\alef^{\rho\pi\pm}\equiv \frac{1}{2}{\rm Arg}\left [e^{-2i \beta} \frac
{A(\ob \to \rho^{\mp} \pi^{\pm})}
{A(B^0 \to \rho^{\pm} \pi^{\mp})}\right ]~,
\eeq
one has
\beq
\alef^{\rho\pi} = \frac{1}{2}(\alef^{\rho\pi+} + \alef^{\rho\pi-})~.
\eeq
The two angle differences $\alef^{\rho\pi\pm} - \alpha$ have
expressions similar to Eq.~(\ref{al1}):
\beq\label{al+-}
\alef^{\rho\pi\pm} - \alpha = \frac{1}{2}\arctan\left [
\frac{2r_{\pm}\sin\gamma(\cos\delta_{\pm} + r_{\pm}\cos\gamma)}
{1 - r_{\pm}^2 + 2r_{\pm}\cos\gamma(\cos\delta_{\pm} + r_{\pm}\cos\gamma)}\right ]~.
\eeq
In the limit $r_{\pm} \to 0$ one obviously obtains $\alef^{\rho\pi\pm}
\to \alpha$.

Flavor SU(3) relates tree and penguin amplitudes in $B\to\rho\pi$ to 
corresponding contributions in $B\to K^*\pi$ and $B\to \rho K$. Introducing  
SU(3) breaking in tree amplitudes in terms of ratios of suitable 
decay constants, and using measured decay rates for the above
processes, the following two upper bounds were obtained at 
90$\%$ CL~\cite{GZ}: $|\alef^{\rho\pi+}-\alpha| < 11^\circ,
|\alef^{\rho\pi-} - \alpha| < 15^\circ$. The algebraic average of
these bounds provides an upper limit on $|\alef^{\rho\pi} -\alpha|$,
\beq\label{bound+}
|\alef^{\rho\pi} -\alpha| < \frac{1}{2}\left (|\alef^{\rho\pi+} -\alpha| + |\alef^{\rho\pi-} -
\alpha|\right ) < 13^\circ~.
\eeq
This bound, which may be modified to $17^\circ$ by SU(3) breaking 
%MG added 
other than in tree amplitudes and by small annihilation amplitudes which
have been neglected, does not assume 
knowledge of the signs of $\alef^{\rho\pi\pm} - \alpha$.  It becomes stronger 
when $\alef^{\rho\pi+} - \alpha$ and $\alef^{\rho\pi-} - \alpha$ have
opposite signs, in which case they cancel each other in
$\alef^{\rho\pi} - \alpha$. This possibility will be discussed now.

The sings of $\alef^{\rho\pi\pm} -\alpha$ depend on values of
$r_{\pm}$ and $\delta_{\pm}$. As we argued in the case of
$B\to\pi^+\pi^-$, both $\alef^{\rho\pi+} -\alpha$ and
$\alef^{\rho\pi-} -\alpha$ would be positive if $r_{\pm} \le 1$ and
$|\delta_{\pm}| \le \pi/2$. Arguments for $r_{\pm} \le 1$ are very
strong. Studies based on flavor SU(3)~\cite{GZ,CGLRS} and a
calculation using QCD-factorization~\cite{BN} obtain rather small
values $r_{\pm} \sim 0.2$, lying very comfortably in the range
$r_{\pm} < 1$.  Both studies obtain values for $\delta_+$ and
$\delta_-$ lying in opposite hemispheres. The output of a global SU(3)
fit is~\cite{CGLRS} $\delta_+ = (178 \pm 14)^\circ$ and $\delta_- = (20\pm
20)^\circ$. Similar values are obtained in~\cite{BN}, where it is
being argued that these values do not differ much from the naive
factorization predictions: $\delta_+=\pi$ and
$\delta_-=0$~\cite{Lipkin}. Using the above values for $r_{\pm}$ and
$\delta_{\pm}$ we see that  $\cos\delta_- + r_- \cos\gamma > 0$
while $\cos\delta_+ + r_+ \cos\gamma < 0$. This implies that 
$0 < \alef^{\rho\pi-} - \alpha < 15^\circ$ but $-11^\circ < \alef^{\rho\pi+} 
- \alpha <0^\circ$, which means that the upper bound (\ref{bound+}) is 
replaced by a stronger bound,
\beq\label{bound-}
-6^\circ < \alef^{\rho\pi} - \alpha< 8^\circ~.
\eeq
That is, given $r_{\pm} \ll 1$, and assuming that the phases
$\delta_+$ and $\delta_-$ lie in opposite hemispheres, improves the
upper bound (\ref{bound+}) by about a factor two. The actual upper and
lower bounds in (\ref{bound-}) may be larger by about 30$\%$ because of
possible SU(3) breaking corrections in penguin amplitudes and small annihilation 
amplitudes which have not been taken into account.

A direct verification of our assumptions about $\delta_{\pm}$ in
$B\to\rho^{\pm}\pi^{\mp}$ relies on interference between tree and
penguin amplitudes in these processes.  A possible evidence for such
interference is the measured 
%MG direct
direct CP asymmetry in
$B^0\to\rho^-\pi^+$~\cite{HFAG,BArhopi,BErhopi},
\beq
A_{\rm CP}(B^0\to\rho^-\pi^+) = -0.48 \pm 0.14~.
\eeq 
This asymmetry provides one equation for $r_-,~\delta_-$ and
$\gamma$~\cite{GZ} favoring negative values of $\delta_-$.  Similarly,
the 
%MG direct
direct CP asymmetry in $B^0\to\rho^+\pi^-$~\cite{HFAG,BArhopi,BErhopi}
\beq
A_{\rm CP}(B^0\to\rho^+\pi^-)=-0.16 \pm 0.09~,
\eeq
provides an equation for $r_+, \delta_+$ and $\gamma$. Two other
observables depending on these four hadronic parameters and on
$\gamma$ are $\bar S$ and $\Delta \bar S$, related to $S$ and $\Delta
S$ by a simple transformation~\cite{GZ}. They obtain 
values~\cite{BArhopi,BErhopi}
\beq 
\bar S = -0.13 \pm
0. 11~,~~~\Delta \bar S = 0.07 \pm 0.11~(0.19)~. 
\eeq 
The error in parentheses includes a rescaling factor of 1.7. 

The dependence of $\bar S$ and $\Delta\bar S$ on a small relative 
phase $\delta_t$~\cite{GZ,CGLRS,BN} between the two tree 
amplitudes $t_+$ and $t_-$ can be neglected. 
%MG added sentence
Alternatively, one may use a constraint from the phase 
${\rm Arg}[A(B^0\to\rho^+\pi^-)A^*(B^0\to\rho^-\pi^+)]$
measured through the interference of the $\rho^+$ and $\rho^-$ 
overlapping resonances in the $B^0\to\pi^+\pi^-\pi^0$ Dalitz
plot~\cite{BArhopi}.
Another constraint on
$r_{\pm},~\delta_{\pm}$ and $\gamma$ is provided by the upper 
bound (\ref{bound+}), where $\alef^{\rho\pi\pm} - \alpha$ 
are given in Eq.~(\ref{al+-}). This completes a system of 
four (or five) equations and one inequality for five parameters 
%MG added
(or six, if we include $\delta_t$). 
An important question is whether these constraints imply that $\delta_+$ 
and $\delta_-$ lie in opposite hemispheres,  thereby leading to 
Eq.~(\ref{bound-}). In any case, the inclusion of these five or six relations 
%MG is
is expected to reduce the uncertainty in $\alpha$ below the upper 
bound (\ref{bound+}).

In conclusion, we have shown that the generic assumptions, $|{\rm
Tree}| \ge |{\rm Penguin}|$ and $|{\rm Arg}({\rm Penguin}/{\rm
Tree})|\le \pi/2$, predict $\alpha < \alef$ in $B\to\pi^+\pi^-$ and
$B\to\rho^+\rho^-$, thereby reducing the theoretical uncertainty in
$\alpha$ by a factor two. In the decay processes
$B\to\rho^{\pm}\pi^{\mp}$, where two pairs of tree and penguin
amplitudes occur, the assumption that the two relative phases between
tree and penguin amplitudes lie in opposite hemispheres results in a
suppression of about a factor two of the bound on $|\alef - \alpha|$
due to cancellation between two terms.  Measured CP asymmetries and
$\alef - \alpha$ were studied in terms of Penguin-to-Tree ratios and
their relative strong phases.  We presented 
%MG added
theoretical arguments, based on QCD-factorization and on flavor SU(3), and 
experimental evidence (although not yet conclusive) that the various hadronic 
parameters do lie in ranges that allow a reduction of the error on $\alpha$.

\bigskip
We wish to thank Ahmed Ali, Andreas Hoecker and Dan Pirjol for helpful
discussions. M. G. is grateful to the Institute for Theoretical
Physics at the University of Zurich for its kind hospitality. This
work is supported in part by the Schweizerischer Nationalfonds and by
the Israel Science Foundation founded by the Israel Academy of
Sciences and Humanities, Grant No. 1052/04.

% Journal definition
\def \ajp#1#2#3{Am.\ J. Phys.\ {\bf#1}, #2 (#3)}
\def \apny#1#2#3{Ann.\ Phys.\ (N.Y.) {\bf#1}, #2 (#3)}
\def \app#1#2#3{Acta Phys.\ Polonica {\bf#1}, #2 (#3)}
\def \arnps#1#2#3{Ann.\ Rev.\ Nucl.\ Part.\ Sci.\ {\bf#1}, #2 (#3)}
\def \art{and references therein}
\def \cmts#1#2#3{Comments on Nucl.\ Part.\ Phys.\ {\bf#1}, #2 (#3)}
\def \cn{Collaboration}
\def \cp89{{\it CP Violation,} edited by C. Jarlskog (World Scientific,Singapore, 1989)}
\def \econf#1#2#3{Electronic Conference Proceedings {\bf#1}, #2 (#3)}
\def \efi{Enrico Fermi Institute Report No.}
\def \epjc#1#2#3{Eur.\ Phys.\ J.\ C {\bf#1} (#3) #2}
\def \ib{{\it ibid.}~}\def \ibj#1#2#3{~{\bf#1} (#3) #2}
\def \ijmpa#1#2#3{Int.\ J.\ Mod.\ Phys.\ A {\bf#1}, #2 (#3)}
\def \ite{{\it et al.}}\def \jhep#1#2#3{JHEP {\bf#1}, #2 (#3)}
\def \jpb#1#2#3{J.\ Phys.\ B {\bf#1}, #2 (#3)}
\def \kdvs#1#2#3{{Kong.\ Danske Vid.\ Selsk., Matt-fys.\ Medd.} {\bf #1}, No.\#2 (#3)}
\def \mpla#1#2#3{Mod.\ Phys.\ Lett.\ A {\bf#1}, #2 (#3)}
\def \nat#1#2#3{Nature {\bf#1}, #2 (#3)}
\def \nc#1#2#3{Nuovo Cim.\ {\bf#1}, #2 (#3)}
\def \nima#1#2#3{Nucl.\ Instr.\ Meth.\ A {\bf#1}, #2 (#3)}
\def \npb#1#2#3{Nucl.\ Phys.\ B~{\bf#1} (#3) #2}
\def \npps#1#2#3{Nucl.\ Phys.\ Proc.\ Suppl.\ {\bf#1}, #2 (#3)}
\def \PDG{Particle Data Group, D. E. Groom \ite, \epjc{15}{1}{2000}}
\def \pisma#1#2#3#4{Pis'ma Zh.\ Eksp.\ Teor.\ Fiz.\ {\bf#1}, #2 (#3) [JETPLett.\ {\bf#1}, #4 (#3)]}
\def \pl#1#2#3{Phys.\ Lett.\ {\bf#1}, #2 (#3)}
\def \pla#1#2#3{Phys.\ Lett.\ A {\bf#1}, #2 (#3)}
\def \plb#1#2#3{Phys.\ Lett.\ B {\bf#1} (#3) #2} 
\def \prd#1#2#3{Phys.\ Rev.\ D\ {\bf#1} (#3)  #2}
\def \prl#1#2#3{Phys.\ Rev.\ Lett.\ {\bf#1} (#3) #2} 
\def \prp#1#2#3{Phys.\ Rep.\ {\bf#1} (#3) #2}
\def \ptp#1#2#3{Prog.\ Theor.\ Phys.\ {\bf#1}, #2 (#3)}
\def \rmp#1#2#3{Rev.\ Mod.\ Phys.\ {\bf#1}, #2 (#3)}
\def \rp#1{~~~~~\ldots\ldots{\rm rp~}{#1}~~~~~}
\def \yaf#1#2#3#4{Yad.\ Fiz.\ {\bf#1}, #2 (#3) [Sov.\ J.\ Nucl.\ Phys.\ {\bf #1}, #4 (#3)]}
\def \zhetf#1#2#3#4#5#6{Zh.\ Eksp.\ Teor.\ Fiz.\ {\bf #1}, #2 (#3) [Sov.\Phys.\ - JETP {\bf #4}, #5 (#6)]}
\def \zp#1#2#3{Zeit.\ Phys.\  {\bf#1} (#3) #2}
\def \zpc#1#2#3{Zeit.\ Phys.\ C {\bf#1}, #2 (#3)}
\def \zpd#1#2#3{Zeit.\ Phys.\ D {\bf#1}, #2 (#3)}

\end{document}